\documentclass[aps,prd,eqsecnum,onecolumn,showpacs,nofootinbib,preprintnumbers,a4paper]{revtex4}

    \newcommand*{\be}{\begin{equation}}
    \newcommand*{\ee}{\end{equation}}
    \newcommand*{\ba}{\begin{eqnarray}}
    \newcommand*{\ea}{\end{eqnarray}}
    
    \newcommand*{\LL}[0]{\mathcal{L}}
    \newcommand*{\PP}[0]{\mathcal{P}}
    \newcommand*{\KK}[0]{\mathcal{K}}
    \newcommand*{\bk}[0]{\bar k}

\usepackage[applemac]{inputenc}
\usepackage[T1]{fontenc} 
\usepackage{comment,cancel,amsmath,amsthm,amssymb,braket,graphicx,pifont,hyperref}
\usepackage{natbib}

\begin{document}

\title{Cosmological Perturbations in the ``Healthy Extension'' of Ho\v{r}ava-Lifshitz gravity}


\author{Alessandro \surname{Cerioni}}
\affiliation{Dipartimento di Fisica, Universit\`a degli Studi di
Bologna, via Irnerio 46, I-40126 Bologna, Italy}
\affiliation{INFN, Sezione di Bologna,
via Irnerio 46, I-40126 Bologna, Italy}
\affiliation{Department of Physics, McGill University, Montr\'eal, QC, H3A 2T8, Canada}
\email{cerioni@bo.infn.it, rhb@physics.mcgill.ca}

\author{Robert H.\ \surname{Brandenberger}}
\affiliation{Department of Physics, McGill University, Montr\'eal, QC, H3A 2T8, Canada}
\email{rhb@mx0.hep.physics.mcgill.ca}

\begin{abstract}
We study linear cosmological perturbations in the ``healthy extension'' of
Ho\v{r}ava-Lifshitz gravity which has recently been analyzed \cite{BPS2}.
We find that there are two degrees of freedom for scalar metric fluctuations,
but that one of them decouples in the infrared limit. Also, for appropriate
choices of the parameters defining the Lagrangian, the extra mode can
be made well-behaved even in the ultraviolet.
\end{abstract}

\pacs{98.80.Cq}

\keywords{Ho\v{r}ava-Lifshitz Gravity, Cosmological Perturbations}

\date{\today}

\maketitle

\section{Introduction}

Following Ho\v{r}ava's proposal \cite{Horava1} of a power-counting
renormalizable Lagrangian for gravity based on anisotropic scaling
between space and time \footnote{See e.g. \cite{Silke,Shinji}
for general overviews.}, several conceptual challenges to the
theory were raised (see e.g. 
\cite{Charmousis,Li,BPS1,Kocharyan,Bogdanos}). One of the
key challenges concerns the extra scalar metric degree of freedom which
arises \cite{Horava1,Cai,Koyama} since there are the same number of variables
appearing in the action, but there is less gauge symmetry. This extra
degree of freedom appears when expanding about flat space-time.
It has been shown \cite{Gao1} that in a particular 
``non-projectable'' version \footnote{A version in which only the terms
in the Lagrangian which appear in the ``projectable'' version are
kept, but in which the lapse function is allowed to depend on space and time.}
of the theory,
the extra mode is non-dynamical in linear cosmological perturbation
theory about an isotropic background (also in the presence of
spatial curvature \cite{Gao2}), although it is expected that the mode
will become dynamical when going beyond linear perturbation theory
\cite{BPS1}. In recent work, we have shown \cite{Cerioni} that 
the extra scalar gravitational mode survives in the ``projectable'' version
of the theory, and that it has either ghost-like or tachyonic properties
\footnote{In pure de Sitter space it appears that this problem is
absent \cite{Wang3}.}.

Fluctuations in Ho\v{r}ava-Lifshitz gravity beyond linear order are
also plagued by a ``strong coupling'' problem \cite{Charmousis,BPS1}.
This problem is absent in the  ``healthy'' extension of the theory 
\cite{BPS2} \footnote{The version of the theory which contains all
the terms in the Lagrangian consistent with the symmetries, 
taking into account that the lapse function now depends on
space and time.}
\footnote{In the far ultraviolet the strong coupling
instability may reappear \cite{Papazoglou}, but at the scale where this
appears higher derivative operators become important and may prevent
these instabilities \cite{BPS3}.}. In recent work \cite{BPS4} it has been
shown that the fluctuations about Minkowski space-time in the
healthy extension of Ho\v{r}ava-Lifshitz (HL) gravity can be made well-behaved
(i.e. neither ghosts nor tachyons) by appropriate choices of the parameters
which enter into the Lagrangian, and that the extra mode does not
lead to any phenomenological inconsistencies. In this note we show
that similar conclusions hold for linear fluctuations about an isotropic
cosmological background: there is a dynamical extra scalar mode, but
it can be made ghost-free and non-tachyonic. In addition, it decouples
in the infrared limit (a limit which cannot be seen when expanding
about Minkowski space-time but a limit which is crucial for actual 
cosmological perturbations) \footnote{We are not the first to consider
cosmological fluctuations in the healthy extension of HL gravity.
In \cite{Yamaguchi2} the evolution of super- and sub-Hubble curvature
fluctuations was studied both analytically and numerically. The
emphasis in our work is on the general properties of the perturbation
modes rather than on the specific form of the solutions of the equations
of motion.}.

The outline of this short paper is as follows: We first give a very brief
review of HL gravity, before writing down the specific Lagrangian of the
healthy extension (Section 2) which will be studied in the following.
In Section 3 we define the linearized cosmological fluctuations. Sections
4 and 5 are the key ones of this work: in Section 4 we derive the
constraint equations and use them to solve for two of the fluctuation
variables. In Section 5 we then derive the quadratic action for the
remaining two cosmological fluctuation variables, making use of
the constraint equations in the derivation. Working in Fourier
space, we then analyze the coefficient matrix of the quadratic
form of the kinetic terms in the action. This yields the conditions
for the absence of ghosts. We study the infrared (IR) and ultraviolet (UV)
limits of the expressions for the eigenvalues of the kinetic coefficient
matrix. We find that in the IR limit one of the eigenvalues tends to
zero. Since the coefficients of the mass matrix do not tend to zero
in the IR limit, the previous result implies that the extra scalar metric
degree of freedom decouples in the IR limit. We discuss the implications
of this result in the concluding section of the paper. In the
penultimate section we make contact with the Minkowski
space-time limit. This is a followup paper to our previous
work \cite{Cerioni} to which the reader is referred for a more
detailed introduction and description of the notation, and for more
references to previous works.

\section{Model}

As in Einstein gravity, in the HL theory the basic variables are the 
metric tensor components in four dimensional space-time. However,
instead of demanding full space-time diffeomorphism invariance,
HL gravity has a reduced set of symmetries - only time-dependent
spatial diffeomorphisms and space-independent time reparametrizations.
Due to the reduced symmetry there is an extra scalar gravitational
degree of freedom.

It is convenient to work in terms of the ADM metric variables, i.e.\
the spatial metric $g_{ij}$, the lapse function $N$ and the shift
vector $N^{i}$ (latin indices run over space components). There
are two versions of HL gravity depending on what coordinates 
$N$ is allowed to depend on. In the projectable version $N$ is
a function of time $t$ only, in the non-projectable version it depends
on both space and time. The ``healthy'' extension studied in \cite{BPS2}
is the version of  the non-projectable model in which all terms
consistent with the residual symmetries are kept, not only those
appearing in the Lagrangian of the projectable version.

The action for HL gravity is based on demanding power-counting
renormalizability under anisotropic scaling between space and time.
We write the action for the healthy extension of HL gravity in the following
form (merging the notations of \cite{Maartens1,Maartens2} and 
\cite{Yamaguchi2}):
\be
S \, = \, \chi^2 \int dt d^3 x N \sqrt{g} \left( \LL_K - \LL_V -\LL_{E}+ \chi^{-2} \LL_M \right)
\ee
where $g\equiv \det(g_{ij})$ and $\chi^2 \equiv 1/(16\pi G)$. The first two terms
in the Lagrangian are present in the initial version of the HL model. They are
the kinetic and potential Lagrangians, respectively. The final term is
the matter Lagrangian. The third term is the new term specific to the
healthy extension \cite{BPS2} of HL gravity. The four terms are given by
\begin{subequations}
\begin{align}
\LL_K &= K_{ij}K^{ij}-\lambda K^2\label{kinetic}\\
\LL_V &= 2\Lambda-R+\frac{1}{\chi^2}(g_2 R^2 +g_3 R_{ij}R^{ij}) + \frac{1}{\chi^4}\left(g_4 R^3 + g_5 R R_{ij} R^{ij} + g_6 R^i_j R^j _k R^k_i \right)+ \\
& +\frac{1}{\chi^4}\left[ g_7 R \nabla^2 R + g_8 (\nabla_i R_{jk})(\nabla^i R^{jk})\right]\nonumber\\
\LL_E &= -\eta a_i a^i + \frac{1}{\chi^2}\left(\eta_2 a_i \Delta a^i + \eta_3 R \nabla_i a^i \right)+ \frac{1}{\chi^4}\left(\eta_4 a_i \Delta^2 a^i + \eta_5 \Delta R \nabla_i a^i +\eta_6 R^2 \nabla_i a^i \right)+...\\
\LL_M &= \frac{1}{2N^2}\left(\dot\varphi-N^i\nabla_i \varphi \right)^2-V(g_{ij},\mathcal{P}_n,\varphi)
\end{align}
\end{subequations}
where $K_{ij}$ is the extrinsic curvature tensor of the constant time 
hypersurfaces, $\Lambda$ is the cosmological constant, 
\be
a_i \, \equiv \, \frac{\partial_i N}{N} \, ,
\ee
and
where for simplicity we have taken matter to be a single scalar field 
$\varphi$.
The constants $\eta, \eta_2, \eta_3, \eta_4$, $\eta_5$ and $\eta_6$ are the parameters that
characterize the specific healthy extension \footnote{We have not written down the full list
of extra terms (see \cite{BPS2} for this list). However, the terms we have omitted
all vanish to quadratic order when expanding about a homogeneous and spatially
flat cosmological background.}. 
Note that the constant $\lambda$ equals
$1$ in General Relativity. In HL gravity, $\lambda$ is expected to flow to $1$
in the IR limit.

For consistency, the matter potential energy should also contain all terms compatible
with the anisotropic scaling symmetry and power-counting renormalizability.  Thus
\cite{Chen}
\be
V \, = \, V_0(\varphi) + V_1(\varphi) \mathcal{P}_0 + V_2(\varphi) \mathcal{P}_1^2 + V_3(\varphi) \PP_1^3 + V_4(\varphi) \PP_2 +V_5(\varphi) \PP_0\PP_2 + V_6(\varphi)\PP_1\PP_2 \, ,
\ee
where
\be
\mathcal{P}_0 \equiv (\nabla\varphi)^2, \, \mathcal{P}_i \equiv \Delta^i \varphi, \, \Delta \equiv g^{ij} \nabla_i \nabla_j \, .
\ee

Note that the only term which may receive contributions from the second order 
$a_i$ in the case of a curved background space-time 
are $R\nabla_i a^i$ and $R^2 \nabla_i a^i$ . Note also
that in Ref.\ \cite{Yamaguchi2}  the $\eta$ coefficients are divided by powers 
of $M_\textrm{Pl}$ instead of powers of $\chi$.

\section{Perturbations}

The scalar metric perturbations can be written as follows (see e.g.
\cite{MFB} for an extensive review article on cosmological fluctuations,
and \cite{Maartens1,Maartens2,Gong} for some other applications of the
theory of cosmological perturbations to HL gravity.):
\begin{subequations}\label{metric_pert}
\begin{align}
\delta N(t,x^k) &= \nu(t,x^k)\\
\delta N_i(t,x^k) &= \partial_i B(t,x^k)\\
\delta g_{ij}(t,x^k) &= a^2(t) \left[-2\,\psi(t,x^k)\,\delta_{ij} + 2\,E(t,x^k)_{|ij}\right] 
\end{align}
\end{subequations}
where the subscript ${}_{|i}$ denotes the covariant derivative.
Correspondingly, also matter fluctuations must be taken into account:
\be\label{matter_pert}
\varphi(t,x^k) = \varphi_0(t) + \delta\varphi(t,x^k)\\
\ee
We can use the freedom under spatial diffeomorphisms to set
\be
\quad E = 0 \, .
\ee
Note that in Ref.\ \cite{Yamaguchi2} the authors use the variable
$\beta \equiv B/a^2$ instead of $B$.

\section{Constraints}

The constraint equations are important since they allow us to reduce the
number of independent dynamical degrees of freedom. We discuss the
Hamiltonian and momentum constraints in turn.

\subsection{Hamiltonian constraint}

The Hamiltonian constraint in the healthy extension of HL gravity takes the form
\be
\LL_K + (\LL_V+\LL_E) + N\frac{\delta (\LL_V + \LL_E)}{\delta N} = \frac{1}{2\chi^2}J^t
\ee
where
\be
J^t \, = \,  2 \bigl( N \frac{\delta {\cal L}_M}{\delta N} + {\cal L}_M \bigr) \, .
\ee
Note that the term 
\be
N\frac{\delta (\LL_V + \LL_E)}{\delta N}  \, = \, 
N\frac{\delta \LL_E}{\delta N} = 2\eta \nabla_i a^i - \frac{2\eta_2 }{\chi^2}\Delta \nabla_i a^i 
+ \frac{\eta_3}{\chi^2} \Delta R - \frac{2\eta_4}{\chi^4} \Delta^2 \nabla_i a^i 
+ \frac{\eta_5}{\chi^4}\Delta^2 R+ \frac{\eta_6}{\chi^6}\Delta R^2 + ...
\ee 
is absent in the ``unhealthy'' HL gravity. Expanding the Hamiltonian constraint to first
order we obtain the following:
\be
\begin{split}
& 3H(3\lambda-1)(\dot\psi+H\nu)+(3\lambda-1)H\Delta B -2\Delta\psi + \frac{\delta\rho_M}{2\chi^2}+ \\ 
& +\eta \Delta\nu  -\frac{\eta_2}{\chi^2}\Delta^2 \nu + 2\frac{\eta_3}{\chi^2}\Delta^2 \psi -\frac{\eta_4}{\chi^4}\Delta^3 \nu + 2\frac{\eta_5}{\chi^4}\Delta^3 \psi+\\
& +3\frac{\KK}{a^2}\left(\frac{\eta_3}{\chi^2}+6\frac{\KK}{a^2}\frac{\eta_6}{\chi^4} \right)\Delta \nu - 6\frac{\KK}{a^2}\left(1-4\frac{\KK}{a^2}\beta_1 -12 \frac{\KK^2}{a^4}\beta_2\right)\psi + \\ 
& +6\frac{\KK}{a^2}(\eta_5 + 4\eta_6 +  2g_7)\frac{\Delta^2\psi}{\chi^4}+ 2\frac{\KK}{a^2}\left(3\frac{\eta_3}{\chi^2}+36\frac{\KK}{a^2}\frac{\eta_6}{\chi^4}+4\beta_1 + 18 \frac{\KK}{a^2}\frac{g_7}{\chi^4}
+ 12\frac{\KK}{a^2}\beta_2\right) \Delta\psi \, = \, 0 \, ,
\end{split}
\ee
where $\KK$ is the spatial curvature constant \footnote{We recall that in the presence
of spatial curvature HL gravity can lead to a non-singular bouncing cosmology
\cite{HLbounce} which is an interesting property of the theory from the point of
view of cosmology.}, 
\be
\beta_1 \equiv \frac{3g_2+g_3}{\chi^2}, \quad \beta_2 \equiv \frac{9g_4+3g_5+g_6}{\chi^4}
\ee
and
\be
\delta\rho_M \, = \,  \dot\varphi_0\dot{\delta\varphi}-\nu \dot \varphi_0^2+ V_{0,\varphi}(\varphi_0) \delta\varphi + V_4(\varphi_0) \Delta^2 \delta\varphi \, .
\ee
Note that the coupling constants $\eta$ newly introduced in the healthy extension 
of HL gravity multiply higher order derivatives of the gravitational degrees of 
freedom, even in the case of a spatially flat background.

\subsection{Momentum constraint}

To first order the momentum constraint in the healthy extension of HL gravity takes 
the form
\be\label{1st_SM_cnstr}
\partial_j\left[(\lambda-1)\Delta B + (3\lambda-1) \left(\dot\psi+H\nu\right) -\frac{2\KK}{a^2}B- \frac{1}{2\chi^2}q_M \right] \, = \, 0 \, ,
\ee
where
\be
q_M \, = \, \dot\varphi_0\delta\varphi \, .
\ee

\subsection{Solving the constraints in a spatially flat background}

Now we specialize to the case of a spatially flat background. Based
on the lessons which can be drawn by comparing cosmological
perturbation theory in the regular non-projectable version of HL
gravity in the cases with \cite{Gao2} and without \cite{Gao1} spatial
curvature, we do not expect any differences with respect to the
number and general properties of dynamical degrees of freedom.

In analogy of what is done in the theory of cosmological perturbations
in the regular non-projectable version of HL gravity \cite{Gao1}, we will
use the constraints to solve for two of the metric degrees of freedom,
namely $\nu$ and $B$. The expressions for $\nu$ and $B$ can be
obtained by combining the two constraint equations. It
is instructive to express the results in terms of the physical
momentum  $\bk \equiv k / a$.  The results can be written in
a compact form if we introduce the notation
\be
f_1(\bk) \, \equiv \, -\eta + \eta_2 \frac{\bk^2}{\chi^2}+\eta_4 \frac{\bk^4}{\chi^4} \, ,
\ee
\be
f_2(\bk) \, \equiv \, -1 + \eta_3 \frac{\bk^2}{\chi^2}+\eta_5 \frac{\bk^4}{\chi^4} \,
\ee
and
\be\label{denominator}
d(\bk) \, \equiv \, 4(3\lambda-1)H^2
+(\lambda-1)\left[\frac{\dot\varphi_0^2}{\chi^2}+ 2f_1(\bk) \bk^2\right] \, .
\ee
We then obtain
\be
\begin{split}
d(\bk)\,B_k(t) =& -(3\lambda-1)\left[\frac{\dot\varphi_0^2}{\chi^2\bk^2}+2 f_1(\bk) \right]\dot \psi_k(t) - (3\lambda-1)\frac{H \dot \varphi_0}{\chi^2\bk^2} \dot{\delta\varphi}_k(t)\\
&-4(3\lambda-1)f_2(\bk) H\psi_k(t) -\Bigg\{(3\lambda-1)[V_{0,\varphi}(\varphi_0) + V_4(\varphi_0)\bk^4]H + \\
&+ 3 (3\lambda-1)\dot\varphi_0 H^2 - \frac{\dot\varphi_0^3}{2\chi^2}-\dot\varphi_0 f_1(\bk) \bk^2\Bigg\}\frac{\delta\varphi_k(t)}{\chi^2 \bk^2}
\\
d(\bk)\,\nu_k(t) = & (\lambda-1)\frac{\dot\varphi_0}{\chi^2}\dot{\delta\varphi}_k(t) - 4(3\lambda-1)H \dot\psi_k(t) +\\
& +  \left\{ (3\lambda-1)\dot\varphi_0 H + (\lambda-1)\left[ V_{0,\varphi}(\varphi_0) + V_4(\varphi_0)\bk^4 \right] \right\} \frac{ \delta\varphi_k(t) }{\chi^2}+\\
& + 4(\lambda-1)f_2(\bk) \bk^2 \psi_k(t) \, .
\end{split}
\ee

Given the form of the common denominator (\ref{denominator}), the solutions 
for $B_k(t)$ and $\nu_k(t)$ are regular in the limit 
$\lambda \rightarrow 1$ whenever $H \neq 0$.

In order to better understand how to interpret the low-momentum and the 
high-momentum limits, it is useful to rewrite the coefficient function $d(\bk)$ of 
Eq.\ (\ref{denominator}) as follows (valid for $H\neq0$ and $\lambda \neq 1/3$):
\be
d(\bk) \, = \, 4(3\lambda-1)H^2  
\left[1+ \frac{\lambda-1}{2(3\lambda-1)}\frac{\dot\varphi_0^2}{\chi^2 H^2}+ \frac{\lambda-1}{2(3\lambda-1)}\left(-\eta + \eta_2 \frac{\bk^2}{\chi^2}+\eta_4 \frac{\bk^4}{\chi^4} \right)\frac{\bk^2}{H^2} \right] \, .
\ee
From this form of the expression, we see that the value of $\bk = k /a$ which separates
the low momentum from the high momentum region is the Hubble momentum $H$. This
is not surprising since we expect fluctuations to behave differently for wavelengths
larger and smaller than the Hubble radius. Note that in the short wavelength 
region $k > aH$, the next-to-leading order terms in the expression for
$d(\bk)$  are controlled by the ratio $k/\chi$ (as long as $H < \chi$ which will
hold in the region of validity of the effective field theory).

In the long wavelength (IR) limit, the expression for $d(\bk)$ becomes
\ba
d(\bk) \, &\sim& \, 4(3\lambda-1)H^2 + (\lambda-1)\dot\varphi_0^2 /\chi^2 \\
 &=& \,  \frac{3\lambda-1}{3}\frac{\dot\varphi_0^2}{\chi^2} 
 + \frac{4}{3}\frac{V_0(\varphi_0)}{\chi^2} + \frac{8}{3}\Lambda \, . \nonumber
\ea
We see that the sign of the first term changes when $\lambda$ crosses the critical value
$\lambda = 1/3$. In the IR limit the first term dominates and hence $d(\bk)$ is
positive. More generally, sufficient conditions for positivity of $d(\bk)$ are
$\lambda > 1/3$, $V_0(\varphi_0)>0$ and $ \Lambda>0$.

In the short wavelength (UV) limit, the expression for $d(\bk)$ becomes
\be
d(\bk) \, \stackrel{\bk \rightarrow \infty}{\longrightarrow} 2(\lambda-1)f_1(\bk) \bk^2 \, ,
\ee
which changes sign as $\lambda$ crosses the value $\lambda = 1$.
 
\section{Second order action}

\subsection{The action}

We are now ready to discuss the second order action for cosmological
fluctuations. We insert the metric ansatz including fluctuations into
the full action, make use of the constraint equations to eliminate the
variables $\nu$ and $B$, and expand. Working in Fourier space, 
after a lot of algebra the terms in the
total action which are second order in the perturbation variables are
\be
\begin{split}
\delta_2 S^{(s)} \, = \, \chi^2 \int dt \frac{d^3 k}{(2\pi)^3} a^3 \Bigg\{ & c_\varphi \dot{\delta\varphi}_k^2 + c_\psi \dot\psi_k^2 +  c_{\varphi\psi}\dot \psi_k \dot{\delta\varphi}_k   + f_\varphi \delta\varphi_k \dot{\delta\varphi}_k+ f_\psi \psi_k \dot\psi_k  +\\  & + f_{\varphi\psi} \psi_k \dot{\delta\varphi}_k +  \tilde f_{\varphi\psi} \dot\psi_k \delta\varphi_k  - m_\varphi^2 \delta\varphi_k^2 -m_\psi^2 \psi_k^2 - m_{\varphi\psi}^2 \psi_k \delta\varphi_k\Bigg\} \, .
\end{split}
\ee

We first focus on the coefficients of the kinetic terms since they will
tell us how many dynamical degrees of freedom survive and whether
there are ghosts. These coefficients are given by
\ba
d(\bk)\,c_\varphi &= & 2(3\lambda-1)\frac{H^2}{\chi^2} + (\lambda-1)f_1(\bk) \frac{\bk^2}{\chi^2}\\
d(\bk)\, c_\psi & = &  2(3\lambda-1) \left[\frac{\dot\varphi_0^2}{\chi^2}+2f_1(\bk) \bk^2\right] \\
d(\bk)\, c_{\varphi\psi} &= & 4 ( 3\lambda-1) \frac{H\dot\varphi_0}{\chi^2} \, .
\ea
We will discuss these terms below. First, we give the expressions
for the coefficients of the terms involving one time derivative of a dynamical
variable. They are
\ba
d(\bk)\,f_\varphi  &= & -(3\lambda-1) \frac{\dot\varphi_0^2 H}{\chi^4} - (\lambda-1) \left[V_{0,\varphi}(\varphi_0) + V_4(\varphi_0)\bk^4 \right]  \frac{\dot\varphi_0}{\chi^4}\\
d(\bk)\, f_\psi &= &  -24 (3\lambda-1) H \Lambda + 12(3\lambda-1)^2 H^3 \nonumber \\
& & - 6\lambda(3\lambda-1) \frac{\dot\varphi_0^2 H}{\chi^2} - 12 (3\lambda-1) \frac{V_0(\varphi_0)H}{\chi^2}\\
d(\bk)\, f_{\varphi\psi} &= &  6(\lambda-1) \frac{\dot\varphi_0 \Lambda}{\chi^2} -3(3\lambda-1)(3\lambda+1) \frac{\dot\varphi_0 H^2}{\chi^2}-\frac{3}{2}(\lambda-1) \frac{\dot\varphi_0^3}{\chi^4} 
\nonumber \\ && + 3 (\lambda-1) \frac{V_0(\varphi_0)\dot\varphi_0}{\chi^4} - 2(\lambda-1)\left[3f_1(\bk) + 2 f_2(\bk)\right] \frac{\dot\varphi_0}{\chi^2}\bk^2\\
d(\bk)\, \tilde f_{\varphi\psi} &= & 4(3\lambda-1) \frac{V_{0,\varphi}(\varphi_0)H}{\chi^2} - (3\lambda-1) \frac{\dot\varphi_0^3}{\chi^4}+\nonumber \\ &&  - 2(3\lambda-1) f_1(\bk) \frac{\dot\varphi_0}{\chi^2}\bk^2 + 4 (3\lambda-1) \frac{V_4(\varphi_0) H}{\chi^2}\bk^4 \, .
\ea
Finally, we also give the expressions for the coefficients of the mass matrix.
They are
\ba
d(\bk)\, m_{\varphi}^2 &=& 2(3\lambda-1) \frac{V_{0,\varphi\varphi}(\varphi_0)H^2}{\chi^2}+\frac{3}{2} (3\lambda-1) \frac{\dot\varphi_0^2 H^2}{\chi^4} + (3\lambda-1) \frac{V_{0,\varphi}(\varphi_0)\dot\varphi_0 H}{\chi^4} + \frac{1}{2} (\lambda-1) \frac{V_{0,\varphi\varphi}(\varphi_0)\dot\varphi_0^2}{\chi^4}+ \nonumber \\  
&& + \frac{1}{2}(\lambda-1)\frac{V_{0,\varphi}(\varphi_0)^2}{\chi^4} - \frac{1}{4}\frac{\dot\varphi_0^4}{\chi^6}   - \Bigg\{4(3\lambda-1) \frac{V_1(\varphi_0) H^2}{\chi^2}- (\lambda-1) f_1(\bk) \frac{V_{0,\varphi\varphi}(\varphi_0)}{\chi^2} +\nonumber \\
&&  +\frac{1}{2}\left[f_1(\bk) + 2(\lambda-1)\right] \frac{\dot\varphi_0^2}{\chi^4} \Bigg\} \bk^2 + \Bigg\{4(3\lambda-1) \left[V_{4,\varphi}(\varphi_0)+V_2(\varphi_0)\right]H^2 - 2(\lambda-1)f_1(\bk) V_1(\varphi_0) + \nonumber \\
&& + (3\lambda-1) \frac{V_4(\varphi_0)\dot\varphi_0 H}{\chi^2} + (\lambda-1) [V_{4,\varphi} (\varphi_0) + V_2(\varphi_0)] \frac{\dot\varphi_0^2}{\chi^2}\Bigg\} \frac{\bk^4}{\chi^2}  + \nonumber \\
&&+ \Bigg\{4(3\lambda-1) V_6(\varphi_0)H^2\chi^2+2(\lambda-1)f_1(\bk) [V_2(\varphi_0) + V_{4,\varphi}(\varphi_0)]\chi^2 + (\lambda-1) V_6(\varphi_0)\dot\varphi_0^2 \Bigg\} \frac{\bk^6}{\chi^4}  + \nonumber \\
&& + \Bigg\{4(\lambda-1) f_1(\bk) V_6(\varphi_0)\chi^4 + \frac{1}{2} (\lambda-1) V_4(\varphi_0)^2\chi^2 \Bigg\} \frac{\bk^8}{\chi^6}
\ea
\ba
d(\bk)\, m_{\varphi \psi}^2 &=& \frac{9}{2} (3\lambda-1)^2 \frac{\dot\varphi_0 H^3}{\chi^2} + \frac{3}{2}(3\lambda-1)(3\lambda-7) \frac{V_{0,\varphi}(\varphi_0)H^2}{\chi^2} - \frac{3}{4} (3\lambda-1)\left[\dot\varphi_0^2 + 2V_0(\varphi_0)\right] \frac{\dot\varphi_0 H}{\chi^4}+\nonumber\\
&& - \frac{3}{2}(\lambda-1) \left[\frac{3}{2}\dot\varphi_0^2 + V_0(\varphi_0)\right] \frac{V_{0,\varphi}(\varphi_0)}{\chi^4} -3(3\lambda-1) \frac{\dot\varphi_0 H \Lambda}{\chi^2} -3(\lambda-1) \frac{V_{0,\varphi}(\varphi_0)\Lambda}{\chi^2} +\nonumber \\ 
&&+  \Bigg\{2(3\lambda-1)f_2(\bk) \frac{\dot\varphi_0 H}{\chi^2} - (\lambda-1) [3f_1(\bk) - 2f_2(\bk)] \frac{V_{0,\varphi}(\varphi_0)}{\chi^2}  \Bigg\}\bk^2-\Bigg\{ 3\frac{V_4(\varphi_0) \Lambda}{\chi^2}+\nonumber\\ &&  - \frac{9}{2}(3\lambda-1) \frac{V_4(\varphi_0)H^2}{\chi^2}+\frac{3}{4} \left[\dot\varphi_0^2 + 2 V_0(\varphi_0) \right]\frac{V_4(\varphi_0}{\chi^4} \Bigg\} (\lambda-1) \bk^4 + 2 (\lambda-1) f_2(\bk) V_4(\varphi_0) \frac{\bk^6}{\chi^2}
\ea
\ba
d(\bk)\, m_{\psi}^2 &=&  3\Lambda \left[4(3\lambda-1) H^2 + (\lambda-1) \frac{\dot\varphi_0^2}{\chi^2} \right]   - \left[\frac{3}{2}(13\lambda-11)\dot\varphi_0^2 - 6V_0(\varphi_0) \right] (3\lambda-1)\frac{H^2}{\chi^2}+\nonumber \\ 
&& -78 (3\lambda-1)^2 H^4 -\frac{3}{4}(\lambda-1) \frac{\dot\varphi_0^4}{\chi^4} + \frac{3}{2} (\lambda-1) V_0(\varphi_0) \frac{\dot\varphi_0^2}{\chi^4} + \Bigg\{6[f_1(\bk) - 4f_2(\bk)] \Lambda+\nonumber \\ 
&& - 3(3\lambda-1) [13f_1(\bk) -12f_2(\bk)] H^2 - \frac{1}{2}[3f_1(\bk) +12f_2(\bk)-4]\frac{\dot\varphi_0^2}{\chi^2} +\nonumber \\ 
&& + 3[f_1(\bk) -4f_2(\bk)] \frac{V_0(\varphi_0)}{\chi^2} \Bigg\} (\lambda-1)\bk^2 + \Bigg\{4(\lambda-1) [f_1(\bk) + 2 f_2(\bk)^2] +\nonumber \\ && +   8 (3\lambda-1) (8g_2 +3g_3) \frac{H^2}{\chi^2} + 2(\lambda-1) (8g_2 + 3g_3) \frac{\dot\varphi_0^2}{\chi^4}\Bigg\}\bk^4 + \Bigg\{4(\lambda-1)f_1(\bk) (8g_2+3g_3)  +\nonumber \\ 
&& + 8(3\lambda-1) (8g_7-3g_8)\frac{H^2}{\chi^2} + 2 (\lambda-1)(8g_7-3g_8)\frac{\dot\varphi_0^2}{\chi^4} \Bigg\}\frac{\bk^6}{\chi^2} + 4(\lambda-1) f_1(\bk) (8g_7-3g_8) \frac{\bk^8}{\chi^4}
 \, .
\ea
Note that the coefficients of the mass matrix remain finite in the limit
$\bk \rightarrow 0$.

\subsection{Observations}

Recall that in the case of the original non-projectable version of HL gravity,
only one of the two metric degrees of freedom was dynamical \cite{Gao1}.
Will this also be the case here? Note that, on setting $\bk = 0$, the kinetic
part of the Lagrangian becomes
\be 
c_\varphi \dot{\delta\varphi}_k^2 + c_\psi \dot\psi_k^2 
+  c_{\varphi\psi}\dot \psi_k \dot{\delta\varphi}_k \, \propto \, 
\left(\frac{H}{\dot\varphi_0}\dot{\delta\varphi}_k+ \dot\psi_k\right)^2 \, .
\ee 
This suggests that the introduction of the Sasaki-Mukhanov variable $\zeta$ defined as
\cite{Sasaki,Mukh}
\be
-\zeta \, \equiv \, \psi + \frac{H}{\dot\varphi_0}\delta\varphi
\ee
may reduce to one the actual number of dynamical degrees of freedom. 

In terms of $\zeta$ and $\psi$, the kinetic part of the Lagrangian is a
quadratic form with coefficients
\ba
d(\bk)\, c_\zeta & =  &\left[2(3\lambda-1) + (\lambda-1) f_1(\bk) \frac{\bk^2}{H^2} \right] \frac{\dot\varphi_0^2}{\chi^2}
\\
d(\bk)\, c_\psi' & = & \left[4(3\lambda-1) + (\lambda-1) \frac{\dot\varphi_0^2}{\chi^2 H^2} \right]f_1(\bk) \bk^2
\\
d(\bk)\, c_{\zeta\psi} & = &
 2 (\lambda-1) \frac{\dot\varphi_0^2}{\chi^2 H^2}f_1(\bk)\bk^2 \, .
\ea
Observe that $c_\psi'$ and $c_{\zeta\psi}$ tend to zero as $\bk \rightarrow 0$, 
whereas $c_\zeta$ is non-trivial as long as the matter field is present.
However,  we see how the presence of the term $f_1(\bk) \neq 0$ - which is present
only in the healthy extension of HL gravity - alters the findings of 
Ref.\ \cite{Gao1}: Both metric degrees of freedom survive as dynamical
variables.

To exclude the possibility that there is a single dynamical metric fluctuation
variable (which would by the above analysis have to be different from
$\zeta$), we need to find the eigenvalues of the coefficient matrix of
the kinetic part of the Lagrangian. Returning to the original variables
$\varphi$ and $\psi$, we consider the kinetic matrix 
\footnote{We have multiplied the $c$'s by proper powers of $\chi$ in order to 
make the matrix dimensionally homogeneous. Such a rescaling is 
equivalent to considering $\delta\varphi_k / \chi$ and $\psi$ as the two 
dynamical variables.}
\be
\left(
\begin{array}{cc}
\chi^2 c_\varphi & \frac{\chi c_{\varphi\psi}}{2}\\
\frac{\chi c_{\varphi\psi}}{2} & c_\psi
\end{array}
\right)
\ee
which has the following eigenvalues:
\be
\begin{split}
& d(\bk) c_{1,2} \, =  \, (3\lambda-1) \left( \frac{\dot\varphi_0^2}{\chi^2}+ H^2 \right) + \frac{13\lambda-5}{2}f_1(\bk) \bk^2 \\
& \pm  \sqrt{ (3\lambda-1)^2 \left (\frac{\dot\varphi_0^2}{\chi^2}+ H^2 \right)^2 + (11\lambda-3)(3\lambda-1) \left( \frac{\dot\varphi_0^2}{\chi^2} - H^2 \right)f_1(\bk) \bk^2 + \left(\frac{11\lambda-3}{2}\right)^2 f_1(\bk)^2 \bk^4}
\end{split}
\ee
It is easy to see that for $f_1(\bk) = 0$ one eigenvalue is exactly zero. However,
in the healthy extension of HL gravity both eigenvalues are non-vanishing and
hence both degrees of freedom are dynamical. This sounds like bad news for
the model. However, we shall now show that in the infrared limit one of the modes
decouples (its mass tends to infinity).

First, however, let us consider under which conditions the linear cosmological
perturbations are ghost-free. We first realize that if
\be\label{eq1}
\frac{1}{d(\bk)}\left[(3\lambda-1) \left( \frac{\dot\varphi_0^2}{\chi^2}+ H^2 \right) + \frac{13\lambda-5}{2}f_1(\bk) \bk^2\right] \, < \, 0 \, ,
\ee
then we will for sure have one negative eigenvalue, i.e. the extra dynamical degree of
freedom will be ghost-like. In the opposite case, in which the expression 
on the left-hand side of Eq.\ (\ref{eq1}) is positive,  one needs to look more
carefully at the expressions. We rewrite $c_{1,2}$ as
\be
c_{1,2} \, = \, A \pm \sqrt{B} 
\ee
and then need to check whether $A^2 - B$ is positive or negative in the case $A>0$. 
The difference is
\be \label{diff}
A^2 - B \, = \, 2(3\lambda-1)\frac{f_1(\bk)\bk^2}{d(\bk)} \, .
\ee
In the IR limit, $f_1(\bk)$ tends to $- \eta$, and thus the condition for ghost freeness
is simply
\be
\eta \, < \, 0 \, .
\ee
For larger values of $\bk$ is not so easy to estimate the sign of the difference
in (\ref{diff}), since we are dealing with several parameters - 
$\lambda,\eta,\eta_2, \eta_4$ - which can have arbitrary signs in the most general 
case. The same difficulty arises when wishing to determine the region in parameter 
space where Eq.\ (\ref{eq1}) is satisfied.

In the following we will assume that the inequality in Eq.\ (\ref{eq1}) is reversed,
and $\eta < 0$ so that the theory is ghost-free in the infrared. First let us consider
the IR limit of the expressions for the eigenvalues $c_{1,2}$. We have
\footnote{We use the approximation $H^2 \gg {\dot \varphi_0}^2 / \chi^2$ which is
justified since the right hand side is one several positive terms which appear
in the expression for $H^2$.}
\ba
c_1 \, & \simeq & \, \frac{1}{2} \, , \\
c_2 \, & \simeq & \, - \eta \bigl( \frac{\bk}{H} \bigr)^2 \, .
\ea
Note that the eigenvalue of the extra degree of freedom goes to zero in the IR
limit. From the expressions for the mass matrix coefficients we see that they
do not tend to zero in the IR limit. Hence, if we re-scale the new scalar gravitational
degree of freedom such that it has canonical kinetic normalization in the IR, we
see that the effective mass of this degree of freedom will diverge as $H / \bk$.
Thus, we see that the extra scalar degree of freedom in the healthy extension of
HL gravity decouples in the IR limit. Hence, at late times, it will not contribute to 
cosmological perturbations on scales relevant to current cosmological observations.
This also explains the results of \cite{Yamaguchi2} who find that the cosmological
fluctuations in the healthy extension of HL gravity agree quite well with those in
Einstein gravity \footnote{There is a potential worry: in terms of the re-scaled variable,
it appears that the theory could be strongly coupled since the coefficients
of the interaction terms will diverge more strongly than the mass (we thank
Omid Saremi for raising this concern). This issue deserves further
study. We believe, however, that there should be no problem since in terms of the original
variables the interaction terms are Planck-suppressed.}. On UV scales, however, 
the extra scalar gravitational mode will
play a role. This may effect the early evolution of fluctuations in inflationary
cosmology on sub-Hubble scales and hence be relevant to the ``trans-Planckian''
problem \cite{TP} of the inflationary universe scenario.

In the UV limit we have 
\be
d(\bk) \, \simeq \, 2(\lambda-1) f_1(\bk) \bk^2 \, ,  
\ee
and from this we can easily show that the eigenvalues are
\ba
c_1 \, & \simeq & \, \frac{1}{2} \, , \\
c_2 \, & \simeq & \, 2 \frac{3 \lambda - 1}{\lambda - 1} \, .
\ea
which is negative for all values of $\lambda$ between $1/3$ and $1$.
In this range of values of $\lambda$ the extra scalar degree of
freedom will be a ghost.

The transition scale between the IR and UV is, as expected, at 
$\bk = H$. Thus, for applications to cosmology the theory should
be ghost-free both in the IR and UV. Hence, we see that the
requirements for a ``good'' behavior of the extra degree of freedom
are $\eta < 0$ and $\lambda > 1$. Note that the
limit $\lambda\rightarrow 1$ is smooth as long as $H\neq 0$. 

\section{Minkowski limit}

Let us finally discuss the Minkowski limit of our analysis which is
obtained if we drop the matter contribution to the action. In this
case, we only have one dynamical degree of freedom for scalar
metric fluctuations, namely $\psi$. We can
observe that $m_\psi^2 = 0$ in the absence of matter and of the 
cosmological constant.

The constraint equations can easily be solved and yield:
\be
\bk^2 B_k(t) \, = \, -\frac{3\lambda-1}{\lambda-1}\dot\psi_k(t)
\ee
and
\be
\nu_k(t) \, = \, \frac{2 f_2(\bk)}{f_1(\bk) }\psi_k(t) \, .
\ee

Inserting these expressions into the action for fluctuations and dropping
the matter terms gives us
\be
\begin{split}
\delta_2 S^{(s)} \, = \, 
2 \chi^2 \int dt \frac{d^3 k}{(2\pi)^3} & \Bigg\{ \frac{3\lambda-1}{\lambda-1}\dot\psi_k^2 
-\left[\left(1+2\frac{f_2(\bk)^2}{f_1(\bk)}\right)\bk^2 + (8g_2+3g_3)\frac{\bk^4}{\chi^2} 
+ (8g_7-3g_8) \frac{\bk^6}{\chi^4 }\right]\psi_k(t)^2 \Bigg\}
\end{split}
\ee

In the IR limit, the equation of motion becomes
\be
\frac{3\lambda-1}{\lambda-1}\ddot\psi_k - \frac{2-\eta}{\eta}\bk^2 \psi_k \, = \, 0 \, .
\ee
We can make the following observations. First, there is
ghost instability for $1/3<\lambda<1$. Secondly, there
is a classical instability unless $0<\eta<2$. These conclusions are
in perfect agreement with those of \cite{Yamaguchi2}.

\section{Discussion and Conclusions}

We have studied cosmological perturbations in the ``healthy'' extension
of HL gravity proposed in \cite{BPS2}, assuming the simplest form
of matter, namely a scalar matter field. Starting from the second order action
for the fluctuations, we studied the general properties of the scalar
modes. We find that there are two dynamical degrees of freedom,
unlike in Einstein gravity and unlike what happens in the non-projectable
version of the original HL model \cite{Gao1}. We identified
the conditions under which the extra degree of freedom in well behaved (i.e.
not a ghost). In the infrared (IR) limit, the condition is $\eta < 0$, where $\eta$ is
one of the constant coefficients defining the extension of the Lagrangian.
In the ultraviolet regime the condition is $\lambda > 1$ or $\lambda < 1/3$.
For the theory to be healthy overall, both the condition on $\lambda$ and
that on $\eta$ have to be satisfied.

Our second main result is that in the IR limit, the mass of the canonically
normalized extra scalar gravitational degree of freedom tends to infinity.
Thus, the extra mode decouples from low energy physics. Hence, when
applied to cosmology, we find that the extra gravitational degree of freedom
is harmless for late-time cosmological perturbations. It may, however,
have interesting consequences for early universe cosmology.

We note that another ``healthy'' extension of HL gravity has recently
been proposed \cite{Horava2}. In that model, the Lagrangian is
constructed so that there is no extra scalar gravitational mode. It would
be interesting to study cosmological fluctuations in that model, as
well.

\begin{acknowledgments}

This work is supported in part by a NSERC Discovery Grant, by funds from
the CRC Program and by a Killam Research Fellowship awarded to R.B. 
We wish to thank Omid Saremi for useful discussions, and P. Ho\v{r}ava and
S. Sibiryakov for helpful comments on the draft of this article.

\end{acknowledgments}

\bibliography{HL}

\begin{thebibliography}{99}

\bibitem{BPS2}
D.~Blas, O.~Pujolas and S.~Sibiryakov,
  ``Consistent Extension Of Horava Gravity,''
  Phys.\ Rev.\ Lett.\  {\bf 104}, 181302 (2010)
  [arXiv:0909.3525 [hep-th]].

\bibitem{Horava1}
P.~Ho\v{r}ava,
  ``Quantum Gravity at a Lifshitz Point,''
  Phys. Rev. {\bf D 79}, 084008 (2009)
 [arXiv:0901.3775 [hep-th]].

\bibitem{Silke}
T.~P.~Sotiriou, M.~Visser and S.~Weinfurtner,
  ``Quantum gravity without Lorentz invariance,''
  JHEP {\bf 0910}, 033 (2009)
  [arXiv:0905.2798 [hep-th]].

\bibitem{Shinji}
S.~Mukohyama,
  ``Horava-Lifshitz Cosmology: A Review,''
  arXiv:1007.5199 [hep-th].

\bibitem{Charmousis}
C.~Charmousis, G.~Niz, A.~Padilla and P.~M.~Saffin,
  ``Strong coupling in Horava gravity,''
  JHEP {\bf 0908}, 070 (2009)
  [arXiv:0905.2579 [hep-th]].
  
\bibitem{Li}
M.~Li and Y.~Pang,
  ``A Trouble with Ho\v{r}ava-Lifshitz Gravity,''
  JHEP {\bf 0908}, 015 (2009)
  [arXiv:0905.2751 [hep-th]].

\bibitem{BPS1}
D.~Blas, O.~Pujolas and S.~Sibiryakov,
  ``On the Extra Mode and Inconsistency of Horava Gravity,''
  JHEP {\bf 0910}, 029 (2009)
  [arXiv:0906.3046 [hep-th]].

\bibitem{Kocharyan}
A.~A.~Kocharyan,
  ``Is nonrelativistic gravity possible?,''
  Phys.\ Rev.\  D {\bf 80}, 024026 (2009)
  [arXiv:0905.4204 [hep-th]].

\bibitem{Bogdanos}
C.~Bogdanos and E.~N.~Saridakis,
  ``Perturbative instabilities in Horava gravity,''
  Class.\ Quant.\ Grav.\  {\bf 27}, 075005 (2010)
  [arXiv:0907.1636 [hep-th]].

\bibitem{Cai}
R.~G.~Cai, B.~Hu and H.~B.~Zhang,
  ``Dynamical Scalar Degree of Freedom in Horava-Lifshitz Gravity,''
  Phys.\ Rev.\  D {\bf 80}, 041501 (2009)
  [arXiv:0905.0255 [hep-th]].

\bibitem{Koyama}
K.~Koyama and F.~Arroja,
  ``Pathological behaviour of the scalar graviton in Ho\v{r}ava-Lifshitz
  gravity,''
  JHEP {\bf 1003}, 061 (2010)
  [arXiv:0910.1998 [hep-th]].

\bibitem{Gao1}
X.~Gao, Y.~Wang, R.~Brandenberger and A.~Riotto,
  ``Cosmological Perturbations in Ho\v{r}ava-Lifshitz Gravity,''
  Phys.\ Rev.\  D {\bf 81}, 083508 (2010)
  [arXiv:0905.3821 [hep-th]].

\bibitem{Gao2}
X.~Gao, Y.~Wang, W.~Xue and R.~Brandenberger,
  ``Fluctuations in a Ho\v{r}ava-Lifshitz Bouncing Cosmology,''
  JCAP {\bf 1002}, 020 (2010)
  [arXiv:0911.3196 [hep-th]].

\bibitem{Cerioni}
A.~Cerioni and R.~H.~Brandenberger,
  ``Cosmological Perturbations in the Projectable Version of Horava-Lifshitz
  Gravity,''
  arXiv:1007.1006 [hep-th].

\bibitem{Wang3}
  Y.~Huang, A.~Wang and Q.~Wu,
  ``Stability of the de Sitter spacetime in Horava-Lifshitz theory,''
  arXiv:1003.2003 [hep-th].

\bibitem{Papazoglou}
A.~Papazoglou and T.~P.~Sotiriou,
  ``Strong coupling in extended Horava-Lifshitz gravity,''
  Phys.\ Lett.\  B {\bf 685}, 197 (2010)
  [arXiv:0911.1299 [hep-th]].

\bibitem{BPS3}
D.~Blas, O.~Pujolas and S.~Sibiryakov,
  ``Comment on `Strong coupling in extended Horava-Lifshitz gravity',''
  Phys.\ Lett.\  B {\bf 688}, 350 (2010)
  [arXiv:0912.0550 [hep-th]].
  
\bibitem{BPS4}
D.~Blas, O.~Pujolas and S.~Sibiryakov,
  ``Models of non-relativistic quantum gravity: the good, the bad and the
  healthy,''
  arXiv:1007.3503 [hep-th].

\bibitem{Yamaguchi2}
T.~Kobayashi, Y.~Urakawa and M.~Yamaguchi,
  ``Cosmological perturbations in a healthy extension of Horava gravity,''
  JCAP {\bf 1004}, 025 (2010)
  [arXiv:1002.3101 [hep-th]].
     
\bibitem{Maartens1}
A.~Wang and R.~Maartens,
  ``Linear perturbations of cosmological models in the Horava-Lifshitz theory
  of gravity without detailed balance,''
  Phys.\ Rev.\  D {\bf 81}, 024009 (2010)
  [arXiv:0907.1748 [hep-th]].

\bibitem{Maartens2}
A.~Wang, D.~Wands and R.~Maartens,
  ``Scalar field perturbations in Horava-Lifshitz cosmology,''
  JCAP {\bf 1003}, 013 (2010)
  [arXiv:0909.5167 [hep-th]].
  

\bibitem{Chen}
 B.~Chen and Q.~G.~Huang,
  ``Field Theory at a Lifshitz Point,''
  Phys.\ Lett.\  B {\bf 683}, 108 (2010)
  [arXiv:0904.4565 [hep-th]].

 \bibitem{MFB}
V.~F.~Mukhanov, H.~A.~Feldman and R.~H.~Brandenberger,
  ``Theory of cosmological perturbations. Part 1. Classical perturbations. Part
  2. Quantum theory of perturbations. Part 3. Extensions,''
  Phys.\ Rept.\  {\bf 215}, 203 (1992).

\bibitem{Gong}
J.~O.~Gong, S.~Koh and M.~Sasaki,
  ``A complete analysis of linear cosmological perturbations in
  Ho\v{r}ava-Lifshitz gravity,''
  Phys.\ Rev.\  D {\bf 81}, 084053 (2010)
  [arXiv:1002.1429 [hep-th]].

\bibitem{HLbounce}
R.~Brandenberger,
   ``Matter Bounce in Horava-Lifshitz Cosmology,''
Phys.\ Rev.\  D {\bf 80}, 043516 (2009)
  [arXiv:0904.2835 [hep-th]].

\bibitem{Sasaki}
M.~Sasaki,
  ``Large Scale Quantum Fluctuations in the Inflationary Universe,''
  Prog.\ Theor.\ Phys.\  {\bf 76}, 1036 (1986).

\bibitem{Mukh}
V.~F.~Mukhanov,
  ``Quantum Theory of Gauge Invariant Cosmological Perturbations,''
  Sov.\ Phys.\ JETP {\bf 67}, 1297 (1988)
  [Zh.\ Eksp.\ Teor.\ Fiz.\  {\bf 94N7}, 1 (1988)].

\bibitem{TP}
   R.~H.~Brandenberger,
  ``Inflationary cosmology: Progress and problems,''
  arXiv:hep-ph/9910410;\\
R.~H.~Brandenberger and J.~Martin, 
``The robustness of inflation to changes in super-Planck-scale physics,"
Mod.~Phys.~Lett.~A~{\bf 16}, 999
(2001), [arXiv:astro-ph/0005432];\\
J.~Martin and R.~H.~Brandenberger,
 ``The trans-Planckian problem of inflationary cosmology,''
Phys.~Rev.~D~{\bf 63}, 123501 (2001), [arXiv:hep-th/0005209].

\bibitem{Horava2}
P. Horava and C. M. Melby-Thompson,
``General covariance in quantum gravity at a Lifshitz point,"
arXiv:1007.2410 [hep-th].
 
\end{thebibliography}

\end{document}